\documentclass[aps,prx,onecolumn,nofootinbib,citeautoscript,10pt]{revtex4-2}  
\usepackage{amsmath,amssymb,bm, makecell}
\usepackage{physics} 

\usepackage{float}

\usepackage{xcolor}
\usepackage{svg}

\usepackage{graphicx}

\usepackage{bbold}

\usepackage[tight]{subfigure} 

\usepackage{color} 
\usepackage[papersize={8.5in,11in}]{geometry}
\usepackage[colorlinks=true]{hyperref}
\hypersetup{        
    unicode=false,          
    pdftoolbar=true,        
    pdfmenubar=true,        
    pdffitwindow=false,     
    pdfstartview={FitH},    
    pdfkeywords={keyword1} {key2} {key3}, 
    pdfnewwindow=true,      
    colorlinks=true,       
    linkcolor=magenta, 
    citecolor=blue,        
    filecolor=magenta,      
    urlcolor=blue           
} 

\usepackage{graphicx}
\usepackage{dcolumn}
\usepackage{color}
\usepackage{amssymb,amsmath}
\usepackage{tabularx,graphicx}
\usepackage{epstopdf}
\usepackage{latexsym}
\usepackage{colortbl}
\usepackage{psfrag}
\usepackage{bbm,bm,array,physics}
\usepackage{dsfont}

 \def\*#1{\mathbf{#1}} 

\newcommand{\be}{\mathbf{e}}

\def\be{\begin{eqnarray}}
\def\ee{\end{eqnarray}}
\def \be{\begin{align}}
\def \ee{\end{align}}
\def \bea{\begin{eqnarray}}
\def \eea{\end{eqnarray}}

\begin{document}

\title{Efficiency of the superconducting diode effect of pair-density-wave states \\in two-dimensional $d$-wave altermagnets}

\author{Igor de M. Froldi}
\affiliation{Instituto de F{\'i}sica, Universidade Federal de Goi{\'a}s, 74.690-900,
Goi{\^a}nia-GO, Brazil}

\author{Hermann Freire}
\affiliation{Instituto de F{\'i}sica, Universidade Federal de Goi{\'a}s, 74.690-900,
Goi{\^a}nia-GO, Brazil}

\begin{abstract}
We systematically study the efficiency of the intrinsic superconducting diode effect of several pair-density-wave states that can emerge in two-dimensional $d$-wave metallic altermagnets. To this end, we investigate several scenarios using an effective minimal microscopic model and Ginzburg-Landau analysis in order to derive the corresponding pairing phase diagrams. In addition, we examine also whether the presence of a Rashba spin-orbit coupling and an applied external magnetic field are beneficial to this effect in these systems. As a consequence, our results add further support to the fact that altermagnetic materials indeed provide a good platform for the pursuit of finite-momentum superconductivity, which can lead to an optimization of the diode efficiency in some physically interesting situations. The latter phenomenon has been recently proposed to be key in improving the applicability of new energy-efficient quantum electronic devices.
\end{abstract}

\maketitle

\section{Introduction} 

Altermagnetism (AM) is a newly discovered type of collinear magnetic phase (see, e.g., Refs. \cite{Smejkal2022_1,Smejkal2022_2,PhysRevX.12.040002})  that shares some properties that are usually linked to ferromagnetism (e.g., time-reversal symmetry breaking, spin-split bands, etc.) and other features normally associated with antiferromagnetism (e.g., zero net magnetization enforced by symmetry). However, the zero net magnetization in AM arises due to a simultaneous application of time-reversal symmetry and a non-trivial rotation that connects the magnetic sublattices (such as, e.g., a four-fold rotation on the square lattice, in the case of a $d$-wave AM), instead of time-reversal and a discrete lattice translation in antiferromagnetism. Consequently, the AM phase breaks Kramer's degeneracy that splits the bands as a result of a new momentum-dependent Zeeman term, which distinguishes AM from the aforementioned conventional magnetic phases. Materials exhibiting AM that were recently confirmed via angle-resolved photoemission (ARPES) experiments include, e.g., MnTe \cite{krempaský_2023,Lee_PRL(2025),Osumi_PRB(2025)} and CrSb \cite{Ding_PRL(2024),Lu_2025,li_2024} (which turn out to be three-dimensional $g$-wave AMs), while strong evidence for $d$-wave AM was observed in quasi-two-dimensional metals KV$_2$Se$_2$O \cite{jiang_arXiv_2024} and Rb$_{1-\delta}$V$_2$Te$_2$O \cite{zhang_arXiv_2024}.

Over the past few years, AM phases have attracted a wide interest in the condensed matter community for several distinct reasons, most notably due to their remarkable applications in the field of spintronics. In this work, we will be especially concerned with the connection regarding the possible emergence of unconventional superconductivity (SC) induced by AM in those materials \cite{Mazin2022notes}. Although the Zeeman splitting of the bands in these systems is generally expected to be detrimental to the formation of conventional $s$-wave singlet superconductivity \cite{Sigrist-RMP(1991)}, other pairing phases that are potentially favored energetically consist of both unconventional singlet and triplet SC states \cite{Mazin2022notes,PhysRevB.108.224421,de_Carvalho_2024,Zhu2023,chakraborty2024constraintssuperconductingpairingaltermagnets,Mukasa_2025,wu2025intraunitcellsingletpairingmediated} and also the elusive pair-density-wave (PDW) phases with Cooper pairs possessing finite center of mass momentum\footnote{Although finite-momentum supercondutivity can manifest itself in terms of several order parameters associated with different broken symmetries (discussed in detail in this work), all these states may be collectively called-pair-density-wave phases. Here, we will use this terminology for conciseness.} \cite{Zhang_Nat_Comm_2024,Chakraborty2024,Knolle-arXiv(2024),Hong_PRB_2025,Hui_Hu_2025}. In this regard, the emergence of superconductivity has recently been found, e.g., in strained thin films of yet another AM candidate material, namely RuO$_2$ \cite{Uchida_PRL_2020}. Despite this observation, the interpretation that this material is a two-dimensional $d$-wave AM has later been challenged by means of muon spin rotation experiments \cite{Hirashi_PRL_2024}. Finally, we point out that AM with a $d$-wave form factor was experimentally shown to emerge using cold atoms in a 2D optical Lieb lattice in the presence of strong Hubbard repulsion \cite{Das_PRL_2024}. The latter systems also provide a good platform for studying unconventional SC and PDW phases induced by AM.

The importance of realizing and experimentally detecting an unambiguous PDW order in quantum materials is a long-sought goal in the field of strongly correlated systems \cite{Agterberg_2020}. Although several recent works
have found compelling evidence for the emergence of such a phase in many important candidate systems, its direct observation has not yet been established beyond doubt. On the theoretical front, many examples of PDW phases have been predicted in various contexts such as, e.g., in effective models in the presence of a large applied magnetic field \cite{FF_paper,osti_4653415}, in strong spin-orbit-coupled systems with broken inversion symmetry \cite{Nagaosa_2018,Yuan_2022}, in many other correlated models using a wide variety of analytical and numerical techniques \cite{Berg2_2009,Loder_2010,Soto_2014,Wang_2015,Freire_2015,Wardh_2017,Carvalho_2021,Wu_2023,Santos_2023,Setty2_2023,Coleman_2022,panigrahi2025microscopictheorypairdensity,Setty_2023,Corboz_2014,Jiang_Kivelson,Xiao_Lee,jiang2023pair,liu2024enhanced,Froldi_2024,zhu2024exact,wang2024pair}, and, quite recently, also in several examples of AM metals in two dimensions (see, e.g., Refs. \cite{Chakraborty2024,Knolle-arXiv(2024),Hong_PRB_2025,Hui_Hu_2025}).

Furthermore, an important motivation for measuring PDW order in strongly correlated systems is that such a phase is expected to lead to the applicability of new energy-efficient quantum electronic devices. For this reason, a distinguishing property that can be associated with some types of PDW states refers to the observation of the superconducting diode effect (SDE) (see, e.g., Refs. \cite{Chakraborty2024,Knolle-arXiv(2024),Yuan_2022,Scheurer-PRB(2024),chakraborty2024perfectsuperconductingdiodeeffect}). This phenomenon refers to the fact that the critical currents in the direction parallel and antiparallel to the Cooper pair momentum are unequal, with the corresponding supercurrents becoming, as a consequence, non-reciprocal (for other recent examples of systems in this context, see, e.g., Refs. \cite{ando2020observation,PhysRevLett.128.037001,PhysRevLett.132.046003,Shaffer_2024,Osin_2024,Scharf_2024,Ili__2024,Qi_2025,Bankier_2025,he2024observationsuperconductingdiodeeffect}, etc.). This effect takes place when both inversion and time-reversal symmetries are broken in the system, due to either an explicit introduction of external magnetic fields or as a result of non-centrosymmetric crystalline environments \cite{Nagaosa_2018,Yuan_2022}, but it can also emerge spontaneously as the property of an emergent SC phase in a given effective model.

To this end, we provide here a systematic study of the emergence of several PDW states, and the efficiency of the SDE that emerges in two-dimensional metallic $d$-wave altermagnets, while also examining the effects of both an applied external magnetic field and the presence of Rashba spin-orbit interaction. Consequently, we establish some interesting scenarios in which this quantity can be enhanced as a function of the physical parameters in the system. In this connection, the main aim of our present study will be to discuss some configurations where optimized values for the SDE efficiency are expected to be observed experimentally, which may have an impact on future applications of this phenomenon in novel candidate AM materials.

This paper is structured as follows. In Sec. II, we define the effective model we are interested in and discuss some preliminary results regarding this system. In Sec. III, we derive microscopically the corresponding Ginzburg-Landau expansion in order to obtain the pairing phase diagram exhibited by this model. Then, in Sec. IV, we obtain the critical currents associated with these phases and compute the corresponding efficiency of the SDE. Finally, in Sec. V, we end with our conclusions and also provide an outlook of some future directions of our present work.

\section{Effective Model}

\begin{figure}[t]
    \centering
    \includegraphics[width=0.7\linewidth]{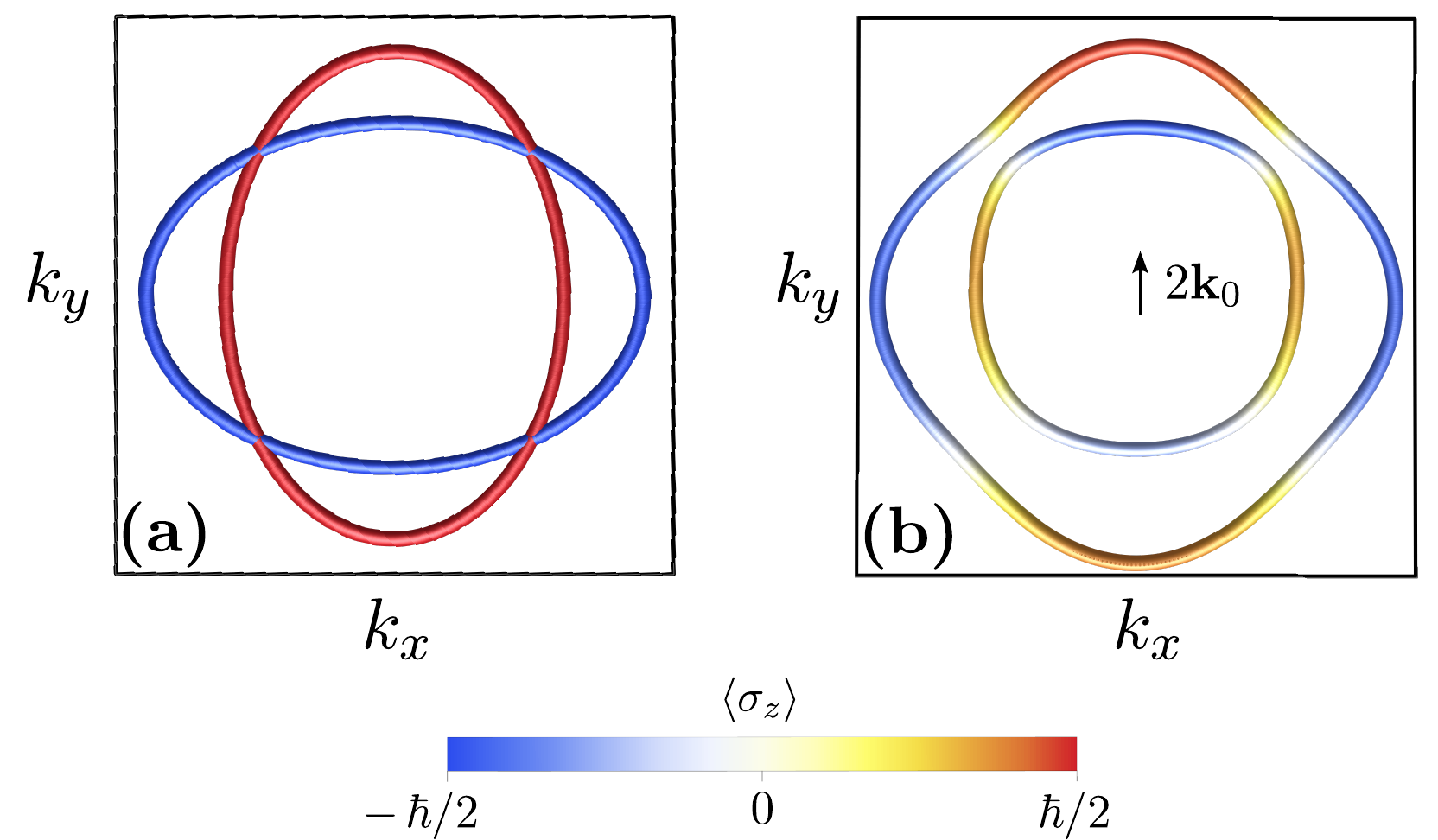}
    \caption{Spin texture of the Fermi surfaces (FS) displayed by the present model described by Eqs. \eqref{FreeH} to \eqref{MagneticH} with (a) $t_{{AM}}=0.7$, $\alpha_{{R}}=\mathbf{B}=0$ and (b) $t_{{AM}}=0.7$, $\alpha_{{R}}=0.6$ and $\mathbf{B}=0.7 \mathbf{\hat{x}}$. As can be seen here, the finite in-plane magnetic field shifts the centers of the inner and outer FS in opposite directions by the vectors $\pm\mathbf{k_0} =\pm
\boldsymbol{\hat{z}}\times \mathbf{B}/v_F$, such that the distance between the two centers is given by $2|\mathbf{k_0}|$. The color of the FS represents the magnitude of the spin polarization $\langle \sigma_z\rangle$.}
    \label{SpinTexture}
\end{figure}

We start with a two-dimensional (2D) $d$-wave AM metal described by the effective Hamiltonian given by $H=H_0+H_{AM}+H_{R}+H_B$, with:
\begin{eqnarray}\label{FreeH}
    H_{0} &=&\sum_{\mathbf{k}, s} \left(\frac{\mathbf{k}^2}{2m} - \mu\right)c^{\dagger}_{\mathbf{k}s}  c_{\mathbf{k}s}\\\label{AM}
    H_{AM} &=& \frac{t_{{AM}}}{2}\sum_{\mathbf{k}, s, s'} c^{\dagger}_{\mathbf{k} s} \left(\boldsymbol{N}_{\mathbf{k}} \cdot \boldsymbol{\sigma} \right)_{ss'} c_{\mathbf{k} s'},\\\label{RashbaH}
    H_{R} &=& \alpha_{R} \sum_{\mathbf{k}, s, s'} c^{\dagger}_{\mathbf{k}s} \left(   \boldsymbol{g}_{\mathbf{k}} \cdot \boldsymbol{\sigma} \right)_{ss'} c_{\mathbf{k}s'},\\\label{MagneticH}
    H_{B} &=& \sum_{\mathbf{k}, s, s'} c^{\dagger}_{\mathbf{k} s} \left( \mathbf{B} \cdot \boldsymbol{\sigma} \right)_{ss'} c_{\mathbf{k} s'},
\end{eqnarray}
where $c^{\dagger}_{\mathbf{k}s}$ ($c_{\mathbf{k}s}$) creates (annihilates) an electron with momentum $\mathbf{k}$ and spin projection $s=\,\uparrow,\downarrow$, $\boldsymbol{\sigma}=(\sigma_x,\sigma_y,\sigma_z)$ is the vector of Pauli matrices,  $\mu$ is the chemical potential, and $m$ stands for the effective mass. The AM spin splitting is parametrized by $t_{AM}$ with the form factor given by $\boldsymbol{N}_{\mathbf{k}} $, which we consider to be in the irreducible representation $B^{-}_{2g}$ of the tetragonal $D_{4h}$ point group (i.e., $\boldsymbol{N}_{\mathbf{k}}\sim(k_x^2-k_y^2)\boldsymbol{\hat{z}}$ in the vincinity of the $\Gamma$ point). Note that $H_{AM}$ breaks the time-reversal symmetry $\mathcal{T}$, since $\boldsymbol{N}_{-\mathbf{k}} = \boldsymbol{N} _{\mathbf{k}}$. Therefore, the Kramer's degeneracy is lifted like a ferromagnet, but with non-uniform splitting set by the $d$-wave form factor, which leads to both spin-polarized bands and zero net magnetization. The momentum-dependent spin-orbit coupling is the Rashba spin-orbit coupling (RSOC) $\alpha_R$, which we will take, as an example, to be described by the form factor given by $\boldsymbol{g}_{\mathbf{k}}= \boldsymbol{\hat{z}} \times \mathbf{k}$ \cite{Yuan_2022}. The RSOC term breaks inversion symmetry $\mathcal{I}$, due to $\boldsymbol{g}_{-\mathbf{k}} = - \boldsymbol{g}_{\mathbf{k}}$, and the interplay between AM and RSOC polarizes the spin degrees of freedom by generating a non-trivial spin texture at the Fermi level for both bands (see Fig. \ref{SpinTexture}). This RSOC can appear, e.g., either by growing the 2D AM metal on a substrate \cite{1984JETPL} or it can emerge in non-centrosymmetric
materials \cite{Agterberg-2004} where $\alpha_R$ naturally arises due to the lack of inversion
of the lattice. We also include here the effects of a uniform magnetic field $\mathbf{B}$ that will be taken to be in-plane, not to affect the AM order parameter which is out-of-plane. Moreover, we will work from now on with natural units ($\hbar=k_B=1$) and, for convenience, we will measure all couplings and energies in units of {$1/(2m)$}. Finally, by following Refs. \cite{Agterberg-2004,Soto_2014}, we normalize the couplings at the non-interacting Fermi surface, such that $\langle \abs{\boldsymbol{g}_{\mathbf{k}}}^2 \rangle_{FS} = \langle \abs{\boldsymbol{N}_{\mathbf{k}}}^2 \rangle_{FS} = 1$, where $\langle \cdot \cdot \cdot \rangle_{FS} = \int_{0}^{2 \pi} (\cdot \cdot \cdot) d \theta/2\pi$.

In order to analyze the emergence of the PDW phases in this model, we will consider here an effective model that has an attractive coupling in the corresponding pairing channel and then we proceed to decouple this quartic interaction using standard mean-field theory. Consequently, we must add to the Hamiltonian described by Eqs. \eqref{FreeH} to \eqref{MagneticH} the following term given by:
\begin{align}\label{HamiltonianSC}
    H_{MF} =& - \sum_{\mathbf{k}. \mathbf{q}}\left(  \gamma(\mathbf{k}) \Delta_{\mathbf{q}} c^{\dagger}_{\mathbf{k} + \frac{\mathbf{q}}{2}, \uparrow } c^{\dagger}_{-\mathbf{k} + \frac{\mathbf{q}}{2},\downarrow} + h.c.\right) +  \sum_{\mathbf{q}} \frac{\abs{\Delta_{\mathbf{q}}}^2}{g},
\end{align}
where $g$ denotes the absolute value of the attractive coupling in the appropriate channel, $\Delta_{\mathbf{q}}\equiv g\sum_{\mathbf{k}}  \gamma(\mathbf{k}) \langle c_{\mathbf{k+q/2},\uparrow}c_{\mathbf{-k+q/2,\downarrow}}\rangle$ is the PDW order parameter, $\mathbf{q}$ is the Cooper pair center-of-mass momentum, and $\gamma(\mathbf{k})$ is a normalized ($\langle \gamma^2(\mathbf{k}) \rangle_{FS}=1$) form factor. We will concentrate here on investigating spin-singlet pairing states for simplicity, and will ignore for the moment the fact that the RSOC (if present) might induce a parity mixing of both singlet and triplet states in the model, as pointed out, e.g., in Refs. \cite{Agterberg-2004,Zhu2023}. Therefore, in what follows, we will analyze the symmetries:
\begin{eqnarray}\label{FormFactor-s-wave}
    \gamma_s &=& 1 \text{\quad ($s$-wave)},\\\label{FormFactor-d-wave}
    \gamma_{d} &=& \frac{\sqrt{2}}{k_F^2} \left(k_x^2 - k_y^2\right) \text{\quad ($d_{x^2-y^2}$-wave)}.
\end{eqnarray}
The modulation of the order parameter also yields different PDW phases, which are given by the following real-space parametrizations \cite{Agterberg_2020,Soto_2014}:

a) \emph{Fulde-Ferrell} (FF) \emph{phase}: 
In this phase, only one wavevector contributes to the order parameter, thereby breaking both time-reversal and inversion symmetries, i.e.
\begin{align}
\Delta(\mathbf{r})=\Delta_{\mathbf{Q}}e^{i\mathbf{Q}\cdot\mathbf{r}}.    
\end{align}

b) \emph{Fulde-Ferrell} (FF*) \emph{phase}: In this phase, the order parameter includes two wavevectors, which are given by $\mathbf{Q}$ and $\widetilde{\mathbf{Q}}=C^{z}_{4} \mathbf{Q}$, with $C^{z}_{4}$ being a fourfold rotation on the square lattice about the $z$-axis, i.e.
\begin{align}
\Delta(\mathbf{r})=\Delta_{\mathbf{Q}}e^{i\mathbf{Q}\cdot\mathbf{r}}+ \Delta_{\mathbf{\widetilde{Q}}}e^{i\mathbf{\widetilde{Q}}\cdot\mathbf{r}}.
\end{align}

c) \emph{Unidirectional} (UD) (or {Larkin–Ovchinnikov}) \emph{PDW phase}: The order-parameter of this phase is a time-reversal invariant version of the FF phase, i.e.  
\begin{align}
\Delta(\mathbf{r})=\Delta_{\mathbf{Q}}e^{i\mathbf{Q}\cdot\mathbf{r}}+\Delta_{-\mathbf{Q}}e^{-i\mathbf{Q}\cdot\mathbf{r}}.
\end{align}

d) \emph{Bidirectional} (BD1) \emph{PDW phase}: The four wavevectors defined previously contribute to the order parameter of this phase, i.e. 
\begin{align}
\Delta(\mathbf{r})=\Delta_{\mathbf{Q}}e^{i\mathbf{Q}\cdot\mathbf{r}}+\Delta_{-\mathbf{Q}}e^{-i\mathbf{Q}\cdot\mathbf{r}}+\Delta_{\mathbf{\widetilde{Q}}}e^{i\mathbf{\widetilde{Q}}\cdot\mathbf{r}}+\Delta_{-\mathbf{\widetilde{Q}}}e^{-i\mathbf{\widetilde{Q}}\cdot\mathbf{r}}.
\end{align}

e) \emph{Bidirectional} (BD2) \emph{PDW phase}: The order parameter of this phase is similar to the previous case, but it now breaks time-reversal symmetry, i.e.  \begin{align}
\Delta(\mathbf{r})=\Delta_{\mathbf{Q}}e^{i\mathbf{Q}\cdot\mathbf{r}}+\Delta_{-\mathbf{Q}}e^{-i\mathbf{Q}\cdot\mathbf{r}}+i(\Delta_{\mathbf{\widetilde{Q}}}e^{i\mathbf{\widetilde{Q}}\cdot\mathbf{r}}+\Delta_{-\mathbf{\widetilde{Q}}}e^{-i\mathbf{\widetilde{Q}}\cdot\mathbf{r}}).
\end{align}

Next, we define the extended Nambu basis given by $\Psi_{ \mathbf{k}} = \left( c_{\mathbf{k}, \uparrow} c_{\mathbf{k}, \downarrow}, c^{\dagger}_{\mathbf{-k}, \uparrow} c^{\dagger}_{\mathbf{-k}, \downarrow} \right)^T$ and write the Hamiltonian in the Bogoliubov-de-Gennes (BdG) form:
\begin{align}\nonumber
    H =& \frac{1}{2} \sum_{\mathbf{k}, \mathbf{k}'} \Psi^{\dagger}_{ \mathbf{k}} \mathcal{\hat{H}}^{\mbox{\scriptsize BdG}}_{\mathbf{k},\mathbf{k}'} \Psi_{ \mathbf{k}'} + \sum_{\mathbf{q}=\pm\mathbf{Q},\pm \mathbf{\widetilde{Q}}} \frac{\abs{\Delta_{\mathbf{q}}}^2}{g},
\end{align}
where {a constant term was neglected since it represents only a shift in the total energy, and $\mathcal{\hat{H}}^{\mbox{\scriptsize BdG}}_{\mathbf{k},\mathbf{k}'} $ is defined as follows}:
\begin{align}\label{BdG}
    \mathcal{\hat{H}}_{\mathbf{k},\mathbf{k}'}^{\mbox{\scriptsize BdG}} =
    \begin{pmatrix}
    \mathcal{\hat{H}}_0(\mathbf{k}) \delta_{\mathbf{k},\mathbf{k}'} & i \sigma_y  \hat{\Delta}(\mathbf{k},\mathbf{k}')  \\
    -i \sigma_y  \hat{\Delta}^{*}(\mathbf{k},\mathbf{k}') &  - \mathcal{\hat{H}}^{T}_0(\mathbf{-k}) \delta_{\mathbf{k}, \mathbf{k}'} 
    \end{pmatrix}.
\end{align}
In Eq. \eqref{BdG}, we used $\mathcal{\hat{H}}_0(\mathbf{k})= \xi_{\mathbf{k}} \sigma_0 + \mathbf{g}^{\mbox{\scriptsize eff}}_ {\mathbf{k}} \cdot \boldsymbol{\sigma}$ and $\hat{\Delta}(\mathbf{k},\mathbf{k}') = \sum_{\mathbf{q} \in \{\pm \mathbf{Q}, \pm \mathbf{\widetilde{Q}}\}} \gamma\left[ \left( \mathbf{k}-\mathbf{k}' \right)/2 \right]\Delta_{\mathbf{q}} \delta_{\mathbf{k}+\mathbf{k}', \mathbf{q}}$, where {the effective coupling is $\mathbf{g}^{\mbox{\scriptsize eff}}_{\mathbf{k}}=\frac{t_{AM}}{2}\boldsymbol{N}_{\mathbf{k}} + \alpha_R\,\boldsymbol{g}_{\mathbf{k}} + \mathbf{B}$}. Then, we obtain the following action from the BdG Hamiltonian:
\begin{align}\nonumber
    \mathcal{S}[\bar{\Psi},\Psi] =& - \sum_{\mathbf{k},\mathbf{k}', i\omega_n} \bar{\Psi}(\mathbf{k},i\omega_n)\left( i \omega_n\mathbb{1} - \mathcal{\hat{H}}^{\mbox{\scriptsize BdG}}_{\mathbf{k}, \mathbf{k}'} \right) \Psi(\mathbf{k}',i\omega_n)\\\label{Action}
    & + \sum_{\mathbf{q}=\{\pm\mathbf{Q} ,\pm \mathbf{\widetilde{Q}}\}} \frac{\beta\abs{\Delta_{\mathbf{q}}}^2}{g},
\end{align}
where $\beta=1/T$ is the inverse temperature in natural units and $\omega_n =(2n+1)\pi T$ are the fermionic Matsubara frequencies ($n\in \mathbb{Z} $). Note that both the BdG Hamiltonian and the identity matrix $\mathbb{1}$ are $4 \times 4$ matrices. Since the BdG Hamiltonian is not diagonal in momentum space, we have to evaluate Eq. \eqref{Action} perturbatively. As we are mainly interested in the vicinity of the SC/PDW transition, we can consider a small $\Delta_{\mathbf{q}}$ in such an expansion. This will be performed in the next section.

\section{Ginzburg-Landau analysis}

To obtain the pairing phase diagram at weak coupling in the model, we derive the corresponding Ginzburg-Landau (GL) free energy. Since the action in Eq. \eqref{Action} is quadratic in the fermionic fields, those degrees of freedom can be integrated out within a path integral representation. Then, we obtain an effective action for the bosonic field $\Delta$, {i.e.}, $\Gamma[\bar{\Delta},\Delta] = \beta \mathcal{F}[\bar{\Delta},\Delta]$, where $\mathcal{F}[\bar{\Delta},\Delta]$ is the free energy. The GL expansion with respect to the order-parameter field up to fourth-order can be written as:
\begin{eqnarray}\label{general_F}
\mathcal{F}&=& \mathit{a}(\mathbf{q}) \sum_{\mathbf{q} =\{\pm\mathbf{Q},\pm\mathbf{\widetilde{Q}}\}} \abs{\Delta_{\mathbf{q}}}^2+\chi_1 \sum_{\mathbf{q} =\{\pm\mathbf{Q},\pm\mathbf{\widetilde{Q}}\}}  \abs{\Delta_{\mathbf{q}}}^4 \nonumber\\ &+& \chi_2 \left( \abs{\Delta_{\mathbf{Q}}}^2 + \abs{\Delta_{-\mathbf{Q}}}^2 \right) \left( \abs{\Delta_{\widetilde{\mathbf{Q}}}}^2 + \abs{\Delta_{-\widetilde{\mathbf{Q}}}}^2 \right)\nonumber\\\label{FourthOrderBD}
    &+&\chi_3 \left( \abs{\Delta_{\mathbf{Q}}}^2 \abs{\Delta_{-\mathbf{Q}}}^2 + \abs{\Delta_{\widetilde{\mathbf{Q}}}}^2 \abs{\Delta_{-\widetilde{\mathbf{Q}}}}^2 \right) \nonumber\\ &+& \chi_4 \left( \Delta_{\mathbf{Q}} \Delta_{-\mathbf{Q}} \Delta^{*}_{\widetilde{\mathbf{Q}}} \Delta^{*}_{-\widetilde{\mathbf{Q}}} + \Delta^{*}_{\mathbf{Q}} \Delta^{*}_{-\mathbf{Q}} \Delta_{\widetilde{\mathbf{Q}}} \Delta_{-\widetilde{\mathbf{Q}}}  \right).\nonumber\\
\end{eqnarray}
The coefficients $\mathit{a}(\mathbf{q})$ and $\chi_i$ were derived from the microscopic Hamiltonian given by Eqs. \eqref{FreeH} to \eqref{MagneticH} and the results obtained are explicitly shown in Appendix \ref{GL}. This form of the GL free energy can be argued solely on a symmetry basis \cite{Agterberg_2008,Agterberg_2020} and was also derived in the context of other microscopic models \cite{Fradkin2007}. {Naturally, the GL approach is only suitable in the vicinity of the critical temperature $T_c$, and the resulting phase diagrams obtained from our analysis should be interpreted as the leading pairing instabilities in the model only within this regime.} For each of the PDW phases explained in the previous section, we calculate the free energy by considering $\abs{\Delta_{\mathbf{q}}} \equiv \Delta$ (with $\mathbf{q} \in \{\pm\mathbf{Q},\pm\mathbf{\widetilde{Q}}\}$), which gives:
\begin{align}\label{free-energy}
\mathcal{F}_{\text{FF}}&=\mathit{a} \abs{\Delta}^2 + \chi_{1} \abs{\Delta}^4,\nonumber\\
\mathcal{F}_{\text{FF*}}&=2 \mathit{a}\abs{\Delta}^2 + \left(2 \chi_1 + \chi_{2} \right) \abs{\Delta}^4,\nonumber\\
\mathcal{F}_{\text{UD}}&=2 \mathit{a} \abs{\Delta}^2 + \left( 2 \chi_1+ \chi_3 \right) \abs{\Delta}^4,\nonumber\\
\mathcal{F}_{\text{BD1}}&= 4\mathit{a} \abs{\Delta}^2 + \left(4 \chi_1 + 4\chi_2 + 2 \chi_3 + 2\chi_4  \right)\abs{\Delta}^4,\nonumber\\
\mathcal{F}_{\text{BD2}}&=4\mathit{a}\abs{\Delta}^2 + \left(4 \chi_1 + 4\chi_2 + 2 \chi_3 - 2\chi_4    \right)\abs{\Delta}^4,
\end{align}
where a constant term in the above expansions was neglected, since it only represents a shift in the free energy. In the above equations, the dependence of $\mathit{a}(\mathbf{q})$ on $\mathbf{q}$ has been omitted to lighten the notation. The coefficients $\mathit{a}$ and $ \chi_i$ describe the stability of the corresponding PDW state with respect to the interactions. By calculating the minimum of the free energy in each PDW phase with respect to the order parameter, we obtain:
\begin{align}\label{FFMin}
    \mathcal{F}^{\text{min}}_{\text{FF}} =& -\frac{\mathit{a}^2}{4\chi_1},\\\label{FFStarMin}
    \mathcal{F}^{\text{min}}_{\text{FF}^*} =&- \frac{\mathit{a}^2}{2\chi_1+\chi_2},\\\label{UDMin}
    \mathcal{F}^{\text{min}}_{\text{UD}} =& - \frac{\mathit{a}^2}{2\chi_1+\chi_3},\\\label{BDMin}
    \mathcal{F}^{\text{min}}_{\text{BD1(2)}} =& -\frac{2 \mathit{a}^2}{2\chi_1 + 2\chi_2 + \chi_3 - \abs{\chi_4}}.
\end{align}
We point out that the difference between the BD1 and BD2 states is given by the sign of the fourth-order coefficient corresponding to $\chi_4<0$ and $\chi_4>0$, respectively. In order to verify which phase is the stable one, we calculate the free energies in Eqs. \eqref{FFMin} to \eqref{BDMin} by considering that the dependence of the fourth-order coefficients on the momentum $\mathbf{q}$ is determined by the momenta $\{ \pm\mathbf{Q},\pm\mathbf{\widetilde{Q}} \}$ that minimize the free energy, giving therefore the stable PDW phase. In case the bidirectional PDW phase corresponds to the true minimum in the GL expansion, we further verify the sign of the corresponding coefficient $\chi_4$ to check if it is a BD of type 1 or type 2. We now study the emergence of such PDW phases in our model for different configurations.

\begin{figure}[t]
     \centering
     \includegraphics[width=0.8\linewidth]{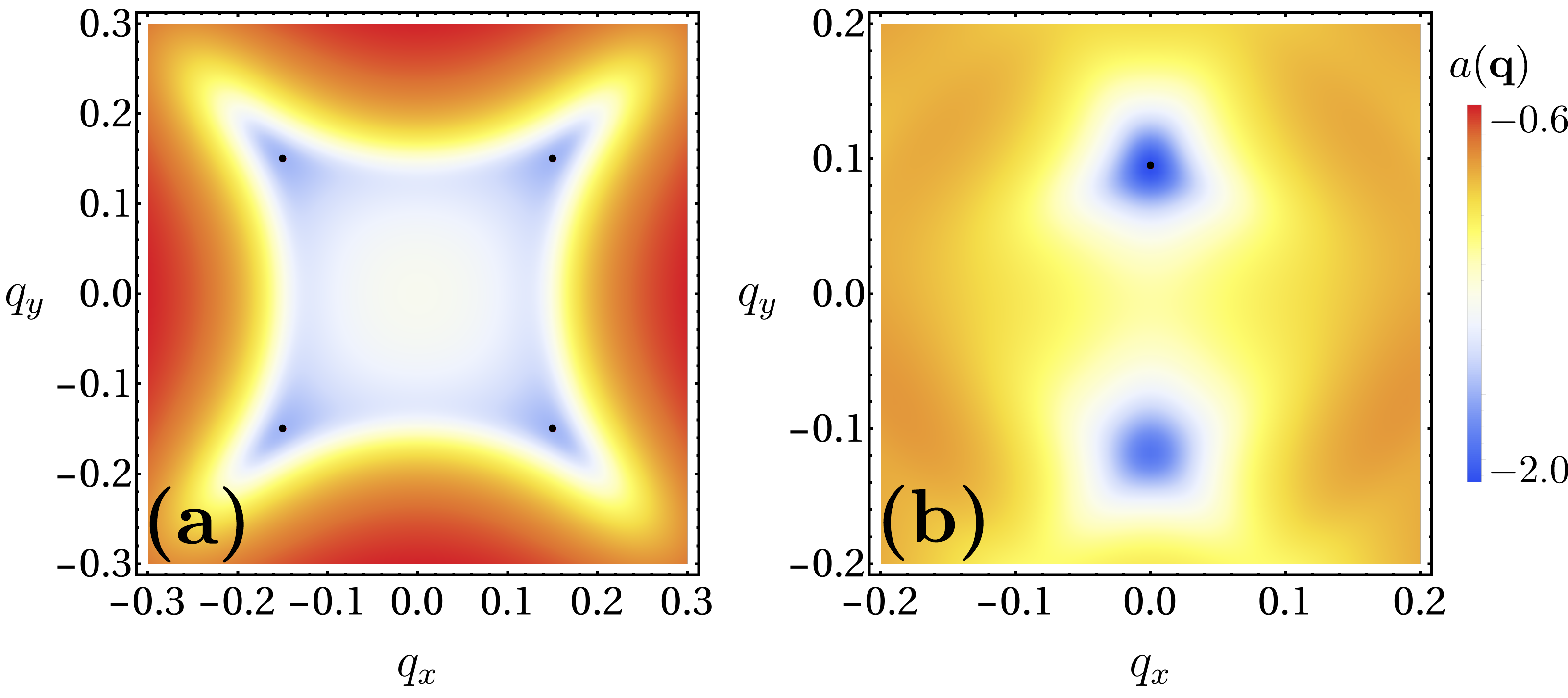}
     \caption{ Second-order coefficient $\mathit{a}(\mathbf{q})$ from Eq. \eqref{free-energy} in the case of an $s$-wave PDW as a function of the momentum $\mathbf{q}$ for {(a)} $t_{AM} = 0.2 $ and $ \alpha_R = \mathbf{B}=0$ and for {(b)} $t_{AM} = \alpha_R = 0.2 $ and $\mathbf{B}\approx 0.9 B_{\text{P}} \mathbf{\hat{x}}$. The black dots indicate the global minima of the free energy, identifying $\mathbf{q}_0$. In \textbf{(a)}, we have $\mathbf{q}_0=(\pm0.15,\pm0.15)$, while in \textbf{(b)} we have only one minimum located at the incommensurate momentum $\mathbf{q}_0=(0.09,0)$ that is proportional to $\mathbf{B}$.}
    \label{Sec_order}
 \end{figure}

\subsection{Altermagnet for $ \alpha_R = \mathbf{B}=0$}

Starting the analysis of the pairing phase diagram described by the GL expansion, we set $\alpha_{\text{R}}=\mathbf{B}=0$, and study the effect of the altermagnetic splitting $t_{{AM}}$, which we shall refer to as the pure altermagnetic case. Although in this case the term set by altermagnetism explicitly breaks time-reversal, inversion symmetry is still preserved at the microscopic level, and the generation of inversion broken states may come only from the spontaneous symmetry breaking in some PDW phases that emerge at low temperatures. With these considerations, the mean-field phase diagram obtained for the present effective model is determined. First, we look for local minima in the second-order GL coefficient $a(\mathbf{q})$ with respect to the momentum $\mathbf{q}$ that is naturally related to the minimum of the free energy. As we increase the altermagnetic splitting, the SC order parameter rapidly acquires a finite Cooper pair momentum $\mathbf{q_0}$. The $d$-wave symmetry of the AM order parameter implies that the momentum associated with such a PDW phase is indeed $\mathbf{q}_0=\{ \pm \mathbf{Q}, \pm \mathbf{\tilde{Q}} \}$, which are connected by fourfold rotational symmetries $C_4^z$ about
the $z$-axis. 
We calculate these minima numerically by discretizing the Brillouin zone (throughout this work, we set $g=1$ for simplicity). The location of $\mathbf{q}_0$ is always found to lie along the directions of the nodal points of the spin-split Fermi surfaces, which can be seen by comparing Figs. \ref{SpinTexture}(a) and \ref{Sec_order}(a). Once confirmed that a PDW phase emerges by the conditions explained above, the fourth-order coefficient in the GL expansion then plays the crucial role of determining which of the symmetrically distinct PDW phases (FF, FF$^*$, UD, BD1 and BD2) minimizes the free energy defined in Eqs. \eqref{FFMin} to \eqref{BDMin}.

We discuss here two possible symmetries for the zero-field PDW that emerges in the present model. For the $s$-wave form factor case [Eq. \eqref{FormFactor-s-wave}], we mention that we only find $\mathcal{I}$-symmetry-preserving PDW states at low temperatures (namely, UD and BD2 phases), in addition to a conventional BCS superconductor and the normal metal state. This indicates that the $s$-wave PDW does not generate non-reciprocal effects due to spontaneous symmetry breaking alone, but in fact needs the application of an external magnetic field and the Rashba spin-orbit interaction, as will be further detailed in the next section.

\begin{figure}[t]
    \centering
    \includegraphics[width=0.45\linewidth]{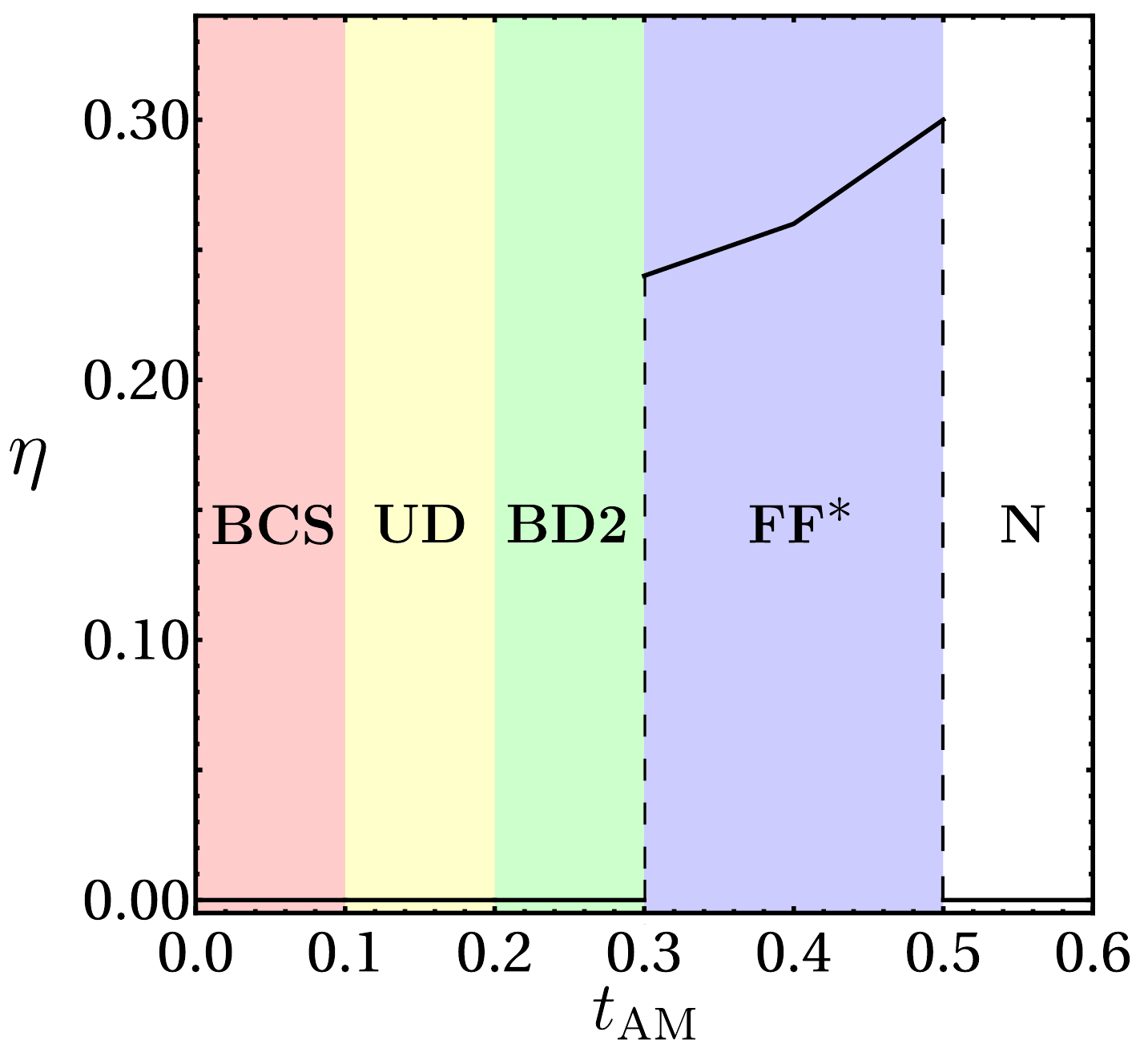}
    \caption{Phase diagram and diode efficiency as a function of the altermagnetic splitting $t_{\text{AM}}$ in the pure altermagnetic case, {i.e.}, $\alpha_{R}=\mathbf{B} = 0$. The ranges of $t_{AM}$ with different colors correspond to different $d_{x^2-y^2}$-wave PDW phases, which are stabilized at low temperatures in the present model. The label ``N'' stands for the normal phase. Note that the SDE efficiency is finite only within the FF$^*$ phase.}
    \label{SDE-tAM}
\end{figure}

By contrast, the corresponding phase diagram becomes much richer for the $d_{x^2-y^2}$-wave PDW case [Eq. \eqref{FormFactor-d-wave}]. As the AM splitting increases, the system transitions from a conventional superconductor to an UD phase ($0.1\lesssim t_{{AM}} \lesssim 0.2$), followed by a BD2 phase ($0.2\lesssim t_{{AM}} \lesssim 0.3$). With further increasing this spin-splitting, a spontaneous break of the inversion symmetry occurs due to the emergence of an FF$^*$ phase ($0.3\lesssim t_{{AM}} \lesssim  0.5$) in the model, and the two selected minima are $\mathbf{Q}$ and $\widetilde{\mathbf{Q}}$. Note that from the stable phases found in this case, only BCS and UD are $\mathcal{T}$-symmetry preserving, which emerge for small $t_{{AM}}$, indicating that a sufficiently large AM splitting is necessary to result in the breaking of time-reversal symmetry of the PDW order parameter itself (see Fig. \ref{SDE-tAM}).

\subsection{Altermagnet with a finite $\alpha_R$ and $\mathbf{B}$}

\begin{figure}[t]
    \centering
    \hspace{-1cm}
    \includegraphics[width=0.55\linewidth]{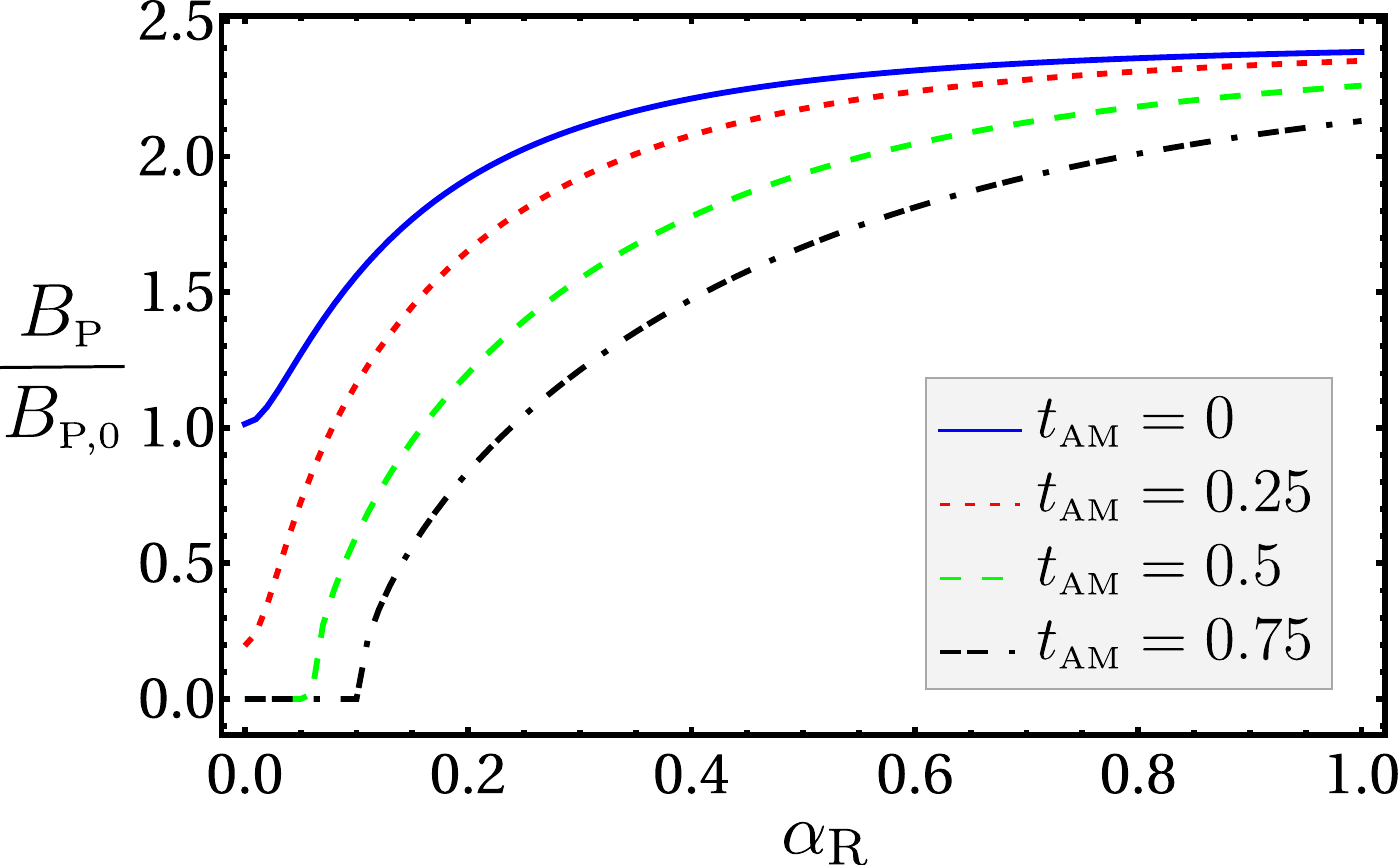}
    \caption{Magnitude of the in-plane Pauli magnetic field along the $x$ direction of the lattice ({i.e.}, $\mathbf{B}_{\text{P}} = B_{\text{P}} \mathbf{\hat{x}}$) normalized by its value $B_{\text{P},0}$ at zero AM splitting, as a function of the RSOC $\alpha_{{R}}$ for different values of $t_{AM}$. Note that, for $\alpha_{{R}}=0$, the Pauli field becomes suppressed as a consequence of the paramagnetic limiting, while  the parameter $\alpha_{{R}}$ increases the Pauli field, even for large AM splitting.}
    \label{BpAlpha}
\end{figure}

We now turn our attention to the situation of a finite applied magnetic field, by focusing on the case of an in-plane field. In this case, we note that for high external magnetic fields, this can cause a transition to the normal state, effectively destroying the singlet superconducting phase. This effect is due to the so-called paramagnetic limiting, and it occurs because $\mathcal{T}$-symmetry breaking terms in the Hamiltonian are detrimental to spin-singlet pairs. Since we are interested here in studying the effects of emergent PDW states in altermagnets as a result of the application of an in-plane field $\mathbf{B}$ [Eq. \eqref{MagneticH}], we also include simultaneously a Rashba spin-orbit coupling $\alpha_{{R}}$ [Eq. \eqref{RashbaH}], as will become clear shortly. The Pauli limiting field $\mathbf{B}_{\text{P}}$ is then calculated as
\begin{align}
\lim_{\mathbf{B} \rightarrow \mathbf{B}_{\text{P}}^-} a(\mathbf{q}) = 0.    
\end{align}
A plot of the Pauli field $\mathbf{B}_{\text{P}}$ as a function of the RSOC coupling for different values of the altermagnetic splitting is given in Fig. \ref{BpAlpha} for the case of the $s$-wave PDW, for simplicity. Note that increasing $t_{{AM}}$ rapidly decreases $\mathbf{B}_{\text{P}}$, since the paramagnetic limiting effect becomes more severe due to the action of both magnetic field and the AM spin-splitting. Interestingly, we find that $\alpha_{{R}}$ causes a growth of $\mathbf{B}_{\text{P}}$, thereby allowing the system in the PDW state to be probed with an applied $\mathbf{B}$ field without destroying it. For this reason, we now study the effect of both $\alpha_R$ and $\mathbf{B}$ in the present model.

Minimizing the GL free energy in the presence of a finite in-plane $\mathbf{B}$ and the Rashba parameter $\alpha_R$, we obtain a conventional BCS superconductor and also a field-induced PDW described by only one momentum $\mathbf{q}_0 = \mathbf{Q}$ (i.e., an FF phase). The generation of such an FF phase can be traced to the fact that the centers of the spin-split Fermi surfaces are shifted by $\mathbf{k}_0 \propto \mathbf{\hat{z}} \times \mathbf{B}/v_F^2$ (where $v_F$ is the Fermi velocity), as seen in Fig. \ref{SpinTexture}(b) for a field along the $x$-axis. The momentum $\mathbf{Q}$ is determined by minimizing the coefficient $a(\mathbf{q})$ in the same way as explained before. Consequently, we find that $\mathbf{Q} \approx 2 \mathbf{k}_0$ for both $s$-wave and $d_{x^2-y^2}$-wave PDW, as shown, e.g., in Fig. \ref{Sec_order}(b) for the case of an $s$-wave form factor, for simplicity. 

In the $s$-wave PDW scenario, we also observe that increasing $t_{AM}$ decreases the magnitude of the momentum $\abs{\mathbf{q}_0}$ of the FF state, while its direction is always fixed along the shifted centers of the spin-split Fermi surfaces. For the $d_{x^2-y^2}$-wave PDW scenario, as the splitting $t_{{AM}}$ is switched on, an FF phase also emerges at low temperatures. For high values of $t_{AM}$, the PDW phase transitions to a normal metal phase in the latter case, as can be seen in the phase diagram of Fig. \ref{SDE-B-tAM}.

Before ending this section, we comment on including only the effect of the RSOC in the model, without the applied in-plane magnetic field. When both $t_{{AM}}$ and $\alpha_{{R}}$ are finite, $\mathcal{T}$ and $\mathcal{I}$ are broken explicitly at the microscopic level. For this reason, one might wonder whether non-reciprocal effects could already be observed in the present model for some appropriate parameters. However, we find that $\alpha_{{R}}$ alone turns out to be detrimental to the formation of spin-singlet PDW states in both $s$ and $d$-wave channels, although the conventional superconducting state with $\mathbf{q}=0$ appears for a larger range of AM splittings than in the $\alpha_{{R}}=0$ case.

\section{Superconducting diode effect}

To calculate the supercurrent from an applied electric field, we use a minimal coupling of the Cooper pair momentum to the field generated by a vector potential $\mathbf{A}$, which reads:
\begin{align}\label{supercurrent}
    \mathbf{J}(\mathbf{q}) = -\frac{\partial \mathcal{F}\left( \mathbf{q} - 2e \mathbf{A} \right)}{\partial \mathbf{A}} \Bigg|_{\mathbf{A} \rightarrow 0}= 2e\frac{\partial \mathcal{F}\left( \mathbf{q}\right)}{\partial \mathbf{q}},
\end{align}
for each of the PDW phases considered in Eqs. \eqref{FFMin} to \eqref{BDMin}. As we explained before, since the SDE emerges when both time-reversal and inversion symmetries are broken in the PDW phase that emerges at low temperatures in the system, the supercurrents that will generate non-reciprocal effects are only those associated with the FF and FF$^*$ states. The other PDW phases (i.e., UD, BD1 and BD2) do not exhibit the diode effect.

\begin{figure}[t]
    \centering
    \hspace{-1cm}
    \includegraphics[width=0.75\linewidth]{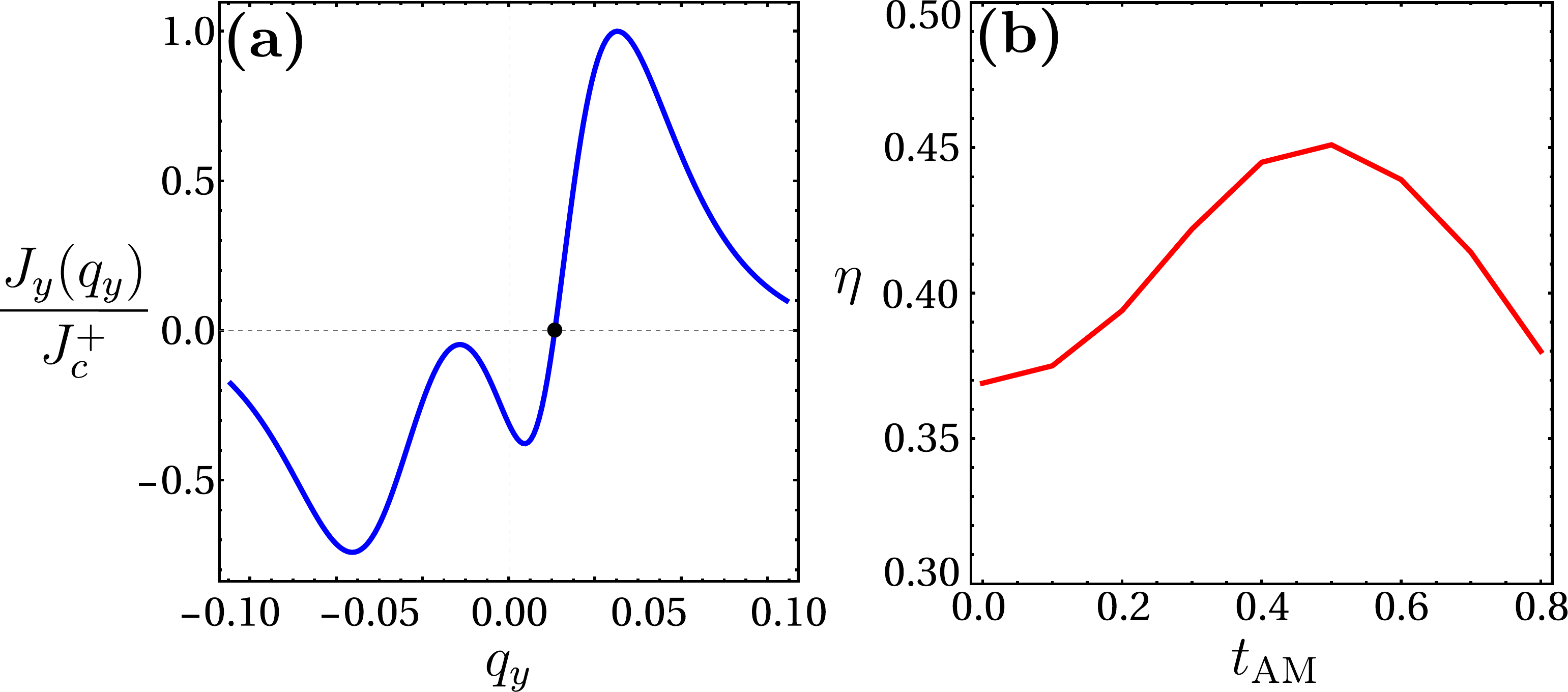}
    \caption{{(a)} Supercurrent as a function of the pair momentum in the FF phase obtained for $t_{{AM}}=0.5, \alpha_{R}=0.2$ and $\mathbf{B} = 0.9 B_{\text{P}}\, \mathbf{\hat{x}}$, within the $s$-wave PDW scenario. For this case, we get the maximum efficiency $\eta^{(s)}_{max} = 0.45$. The black dot indicates the single local minimum ($\mathbf{Q} \approx 0.02\, \mathbf{\hat{y}}$) in the GL free energy. Note that the supercurrent is zero only for the pair momentum  ${\mathbf{Q}}$, and the stable branch is the maximum and minimum region of $\mathbf{J}_{\text{FF}}(\mathbf{q})$ around ${\mathbf{Q}}$. {(b)} Diode efficiency as a function of the splitting $t_{{AM}}$ for $\alpha_{{R}}=0.2$, $\mathbf{B}=0.9B_{\text{P}}\, \mathbf{\hat{x}}$ in the $s$-wave PDW scenario.}
    \label{SDE-B}
\end{figure}

In a conventional BCS superconductor, the supercurrent necessarily vanishes for zero pair momentum, while other zeros of the current are reached only by possible local minima, since the supercurrent is odd with respect to the momentum, i.e., $\mathbf{J}(\mathbf{q})=-\mathbf{J}(-\mathbf{q})$. However, in the PDW case the situation is more complex. For states in which $\mathcal{I}$-symmetry is preserved (namely, the UD, BD1 and BD2 phases) $\mathbf{J}(\mathbf{q})$ is also odd under inversion, but the zeros of the supercurrent along the $\mathbf{q}$-axis correspond to global minima, i.e., the true stable phase, which yields a reciprocal superconductor. For $\mathcal{I}$-symmetry-broken states (FF and FF$^*$ phases), the supercurrent is no longer odd with respect to the momentum $\mathbf{q}$, and two distinct branches emerge: one that is stable corresponding to the global minimum (true low-temperature state) and the other one corresponding to a local minimum (unstable phase). Therefore, in the FF and FF$^*$ phases, an asymmetry is observed between the momenta parallel and antiparallel to the momentum of the Cooper pair $\mathbf{q_0}$ [see, e.g., Fig. \ref{SDE-B}(a)]. This asymmetry is the origin of the so-called supercurrent diode effect.

\begin{figure}[t]
    \centering
    \hspace{-1cm}
    \includegraphics[width=0.45\linewidth]{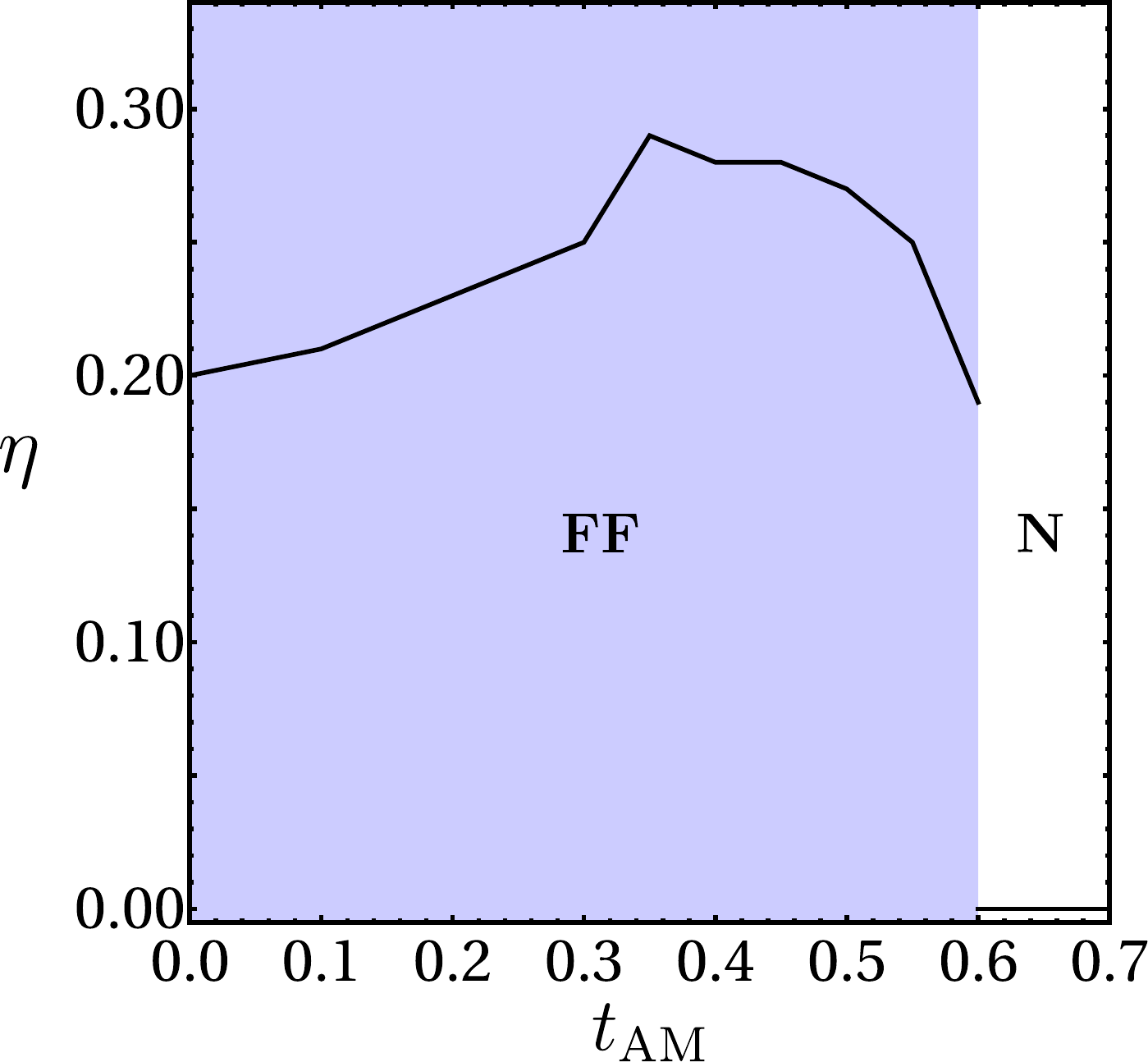}
    \caption{Phase diagram and diode efficiency as a function of  $t_{{AM}}$ for the $d_{x^2-y^2}$-wave PDW scenario for $\alpha_{{R}}=0.2$ and $\mathbf{B} = 0.9 B_{\text{P}} \mathbf{\hat{x}}$. The label ``N'' stands for the normal phase. }
    \label{SDE-B-tAM}
\end{figure}

In order to study the SDE, we focus on the FF and FF$^*$ phases that, as we have seen, have a finite superconducting diode response. Inserting Eqs. \eqref{FFMin} and \eqref{FFStarMin} in Eq. \eqref{supercurrent}, the supercurrents associated with these phases are given, respectively, by:
\begin{align}\label{SuperCurrentFF}
    \mathbf{J}_{\text{FF}}(\mathbf{q})=&e\frac{\abs{a(\mathbf{q})}}{\chi_1} \nabla_{\mathbf{q}} a(\mathbf{q}),\\\label{SuperCurrentFF*}
    \mathbf{J}_{\text{FF}^*}(\mathbf{q})=&4e\frac{\abs{a(\mathbf{q})}}{2\chi_1 +\chi_2} \nabla_{\mathbf{q}} a(\mathbf{q}),
\end{align}
for $a(\mathbf{q})<0$, $\chi_1 >0$ and $2\chi_1+\chi_2>0$. Note that the supercurrents vanish at the equilibrium momentum $\mathbf{q}_0$ due to the gradient term in Eqs. \eqref{SuperCurrentFF} and \eqref{SuperCurrentFF*}, in agreement with Bloch's theorem. 

The efficiency of the SDE that quantifies the non-reciprocal character of the supercurrent exhibited by the corresponding PDW phase is defined as \cite{Yuan_2022}:
\begin{align}\label{Efficiency}
    \eta = \frac{J_c^+-J_c^-}{J_c^++J_c^-},
\end{align}
where $J_c^+$ and $J_c^-$ are, respectively, the maximum and minimum of $\mathbf{J}(\mathbf{q})$ along the direction(s) of $\mathbf{q}_0$ corresponding to the stable FF or FF$^*$ order parameters. 

In the pure altermagnetic case ($\alpha_{{R}} = \mathbf{B}=0$), the only way to get a non-zero SDE occurs when the inversion symmetry is broken spontaneously in the PDW phase. In Fig. \ref{SDE-tAM}, we show for the case of the $d_{x^2-y^2}$-wave PDW scenario that, by varying the AM splitting, we obtain a cascade of first-order phase transitions (BCS $\rightarrow$ UD $\rightarrow$ BD2 $\rightarrow$ FF$^*$), and the diode effect measure $\eta$ is naturally non-zero only when the FF$^*$ state is found. Interestingly, the window for this $\mathcal{I}$-broken state turns out to be larger than the other PDW states (i.e., the BCS, UD and BD2 orders) in the phase diagram. Moreover, we found a maximum efficiency in the field-free FF$^*$ phase  of approximately $\eta^{(d)}_{max} \sim 0.3$, near the second-order phase transition of such a phase into the normal state.

For the case of finite $\alpha_{{R}}$ and in-plane $\mathbf{B}$, we obtain the phase diagrams shown in Figs. \ref{SDE-B-tAM} and \ref{SDE-B}(b). We discuss the efficiency of the SDE in both field-induced $s$-wave and $d_{x^2-y^2}$-wave PDW scenarios. In the first case (Fig. \ref{SDE-B}), as discussed previously, only an FF phase is found for not-too-large AM splittings. Nevertheless, we find that as $t_{AM}$ is increased, the diode efficiency is first slightly enhanced and then, after an optimal point around $t^{(s)}_{AM}\sim 0.5$, it starts to be suppressed\footnote{Here, we point out that, for even higher AM spin-splitting (i.e., $t_{AM}>0.8$), the free energy $\mathcal{F}$ of the model does not exhibit a global minimum any longer and, for this reason, one has to include higher-order terms in the GL expansion. However, the latter study is beyond the scope of the present work.}. Interestingly, we obtain the maximum SDE efficiency of the model ($\eta^{(s)}_{max}\sim 0.45$) around this optimal value, within this scenario. By contrast, for the case of a $d_{x^2-y^2}$-wave PDW phase (Fig. \ref{SDE-B-tAM}), we observe that the maximum diode efficiency is once more approximately $\eta^{(d)}_{max}\sim 0.3$, which occurs near a smaller optimal value ($t^{(d)}_{AM}\sim 0.35$) compared to the $s$-wave PDW scenario.

\section{Conclusions and Outlook}

In this paper, we have systematically studied the emergence of PDW states in $d$-wave altermagnets and the efficiency of the corresponding superconducting diode effect, using a minimal microscopic effective model and Ginzburg-Landau analysis. As a consequence, we have explored the richness of the pairing phase diagrams exhibited by these systems and analyzed, in particular, some distinct signatures of different emergent PDW phases that appear at low temperatures. 

In some interesting scenarios, we have shown that some PDW phases lead to remarkable non-recriprocal transport properties. The underlying mechanism of producing a finite SDE 
in the present model corresponds to breaking both inversion and time-reversal symmetries. Here, we have discussed that this can be achieved in two different ways in altermagnetic systems: either by breaking explicitly these two symmetries at the microscopic level or as a result of a spontaneous symmetry breaking produced by the PDW phase that emerges at low temperatures. In the first case, we have shown that the system describes a field-induced FF phase for a wide range of AM splittings, whereas in the second case obtaining a finite SDE requires fine-tuning of the physical parameters in order to reach a field-free FF$^*$ phase that emerges at low temperatures. Moreover, the maximum efficiency in the first scenario ($\eta^{(s)}_{max}\sim 0.45$) is also higher than in the second scenario ($\eta^{(d)}_{max}\sim 0.3$), which suggests that breaking $\mathcal{I}$ and $\mathcal{T}$ symmetries explicitly in those systems may be more fruitful in order to achieve a high SDE that can be ultimately explored in technological applications.

Finally, we point out that in systems with broken $\mathcal{I}$-symmetry ($\alpha_R\neq 0$), spin-singlet and spin-triplet finite-momentum SC states are expected to be mixed. Therefore, it is worthwhile to study parity-mixed PDW phases emerging in the present model and also in other systems where the AM order parameter transforms under different irreducible representations of other crystallographic point-groups. In the latter case, we have in mind particularly the wide variety of three-dimensional AM models in the presence of spin-orbit coupling which have been recently a focus of research (such as, e.g., the effective model studied in Refs. \cite{de_Carvalho_2024,Fernandes-PRB(2024)}). The goal is to examine whether it is possible to obtain an even higher SDE efficiency in PDW phases that potentially emerge in these systems as a function of the physical parameters. For this reason, we plan to perform these analyses in future publications.

\section*{Acknowledgments}

We would like to thank V. S. de Carvalho for discussions. H.F. acknowledges funding from the Conselho Nacional de Desenvolvimento Cient\'{i}fico e Tecnol\'{o}gico (CNPq) under grant numbers 311428/2021-5, 404274/2023-4 {and 305575/2025-2}.

\appendix

\section{Derivation of the Ginzburg-Landau coefficients from the microscopic model}\label{GL}

After integrating out the fermionic degrees of freedom in Eq. \eqref{Action}, the effective action is given by:
\begin{align}
    \Gamma(\Delta,\bar{\Delta}) =&  \beta \sum_{\mathbf{q} \in \{ \pm \mathbf{Q},\pm \mathbf{\widetilde{Q}} \}}\frac{\abs{\Delta_{\mathbf{q}}}^2}{g} - \Tr \ln \left( \mathcal{\hat{G}}_0^{-1} + \hat{\Delta} \right),
\end{align}    
where:
\begin{align}\label{Propagator}
    \mathcal{\hat{G}}_0^{-1}(\mathbf{k}) =&\begin{pmatrix}
     i\omega_n -\mathcal{H}_0(\mathbf{k})  & 0  \\
    0 &  i\omega_n + \mathcal{H}^{T}_0(\mathbf{-k})  
    \end{pmatrix}  \,
\end{align}
and
\begin{align}
        \hat{\Delta}(\mathbf{k}, \mathbf{k}')=&\begin{pmatrix}
    0 & i \sigma_y  \Delta(\mathbf{k},\mathbf{k}')  \\
    -i \sigma_y  \bar{\Delta}(\mathbf{k},\mathbf{k}') & 0
    \end{pmatrix}
\end{align}
are the inverse free propagator and the order parameter matrices, respectively. Since $\hat{\Delta}(\mathbf{k}, \mathbf{k}') = \sum_{\mathbf{q}} \gamma\left[ \left(\mathbf{k}- \mathbf{k}' \right)/2 \right]\Delta_{\mathbf{q}} \delta_{\mathbf{k}+\mathbf{k}',\mathbf{q}}$, we calculate the free energy in terms of an expansion in the order parameter $\Delta_{\mathbf{q}}$ given by:
\begin{align}\nonumber
   \mathcal{F}\left( \Delta, \bar{\Delta} \right) =&  \sum_{\mathbf{q} \in \{ \pm \mathbf{Q},\pm \mathbf{\widetilde{Q}} \}}\frac{\abs{\Delta_{\mathbf{q}}}^2}{g} - \Tr \ln\left( \mathcal{\hat{G}}_0^{-1} \right) \\\label{EffAcion}
   &+ \frac{1}{\beta}\sum_{n=1}^{\infty} \frac{1}{2n} \Tr \left( \mathcal{\hat{G}}_0  \hat{\Delta} \right)^{2n}.
\end{align}
Each term $n$ in the sum yields a $2n$-th order in the bosonic field $\Delta$, corresponding to a one-particle irreducible bubble diagram associated with $2n$ order-parameter external legs. Notice that the corresponding diagrams with an odd number of the order-parameter legs vanish, so we have neglected them. Before we start the calculation of the quadratic and quartic coefficients in the Ginzburg-Landau free energy, let us rewrite the particle and hole propagators that appear in Eq. \eqref{Propagator}, which are given by:
\begin{align}\nonumber
    G_0^{(p)} \left( \mathbf{k}, i\omega_n \right) 
    =& \frac{1}{i\omega_n-\mathcal{H}_0(\mathbf{k})}\nonumber\\
    =& \frac{\left( -i \omega_n + \xi_{\mathbf{k}} \right) \sigma_0 - \mathbf{g}^{\mbox{\scriptsize eff}}_{\mathbf{k}} \cdot \boldsymbol{\sigma}}{\left( -i \omega_n + \xi_{\mathbf{k}} \right)^2 - \abs{\mathbf{g}^{\mbox{\scriptsize eff}}_{\mathbf{k}}}^2},\\\nonumber
    G_0^{(h)} \left( \mathbf{k}, i\omega_n \right) 
    =& \frac{1}{i\omega_n+\mathcal{H}^T_0(-\mathbf{k})}\nonumber\\
    =& \frac{\left( -i \omega_n - \xi_{\mathbf{-k}} \right) \sigma_0 - \mathbf{g}^{\mbox{\scriptsize eff}}_{-\mathbf{k}} \cdot \sigma_y \boldsymbol{\sigma} \sigma_y}{\left( -i \omega_n - \xi_{\mathbf{-k}} \right)^2 - \abs{\mathbf{g}^{\mbox{\scriptsize eff}}_{-\mathbf{k}}}^2},
\end{align}
in terms of the propagators $G^{(+)}$ and $G^{(-)}$:
\begin{align}
    G^{(+)} \left( \mathbf{k}, i\omega_n \right) =& \left( \frac{1/2}{-i \omega_n + \xi_{\mathbf{k}}+ \abs{\mathbf{g}^{\mbox{\scriptsize eff}}_{\mathbf{k}}}} + \frac{1/2}{-i \omega_n + \xi_{\mathbf{k}} - \abs{\mathbf{g}^{\mbox{\scriptsize eff}}_{\mathbf{k}}}} \right),\\
    G^{(-)} \left( \mathbf{k}, i\omega_n \right) =&  \left( \frac{1/2}{-i \omega_n + \xi_{\mathbf{k}} + \abs{\mathbf{g}^{\mbox{\scriptsize eff}}_{\mathbf{k}}}} - \frac{1/2}{-i \omega_n + \xi_{\mathbf{k}} - \abs{\mathbf{g}^{\mbox{\scriptsize eff}}_{\mathbf{k}}}} \right).
\end{align}
Therefore, we have:
\begin{align}\label{ParticlePropagator}
    G^{(p)}_0 \left(\mathbf{k}, i \omega_n \right) =& G^{(+)} \left( \mathbf{k}, i\omega_n \right) \sigma_0 + G^{(-)} \left( \mathbf{k}, i\omega_n \right) \mathbf{g}^{\mbox{\scriptsize eff}}_{\mathbf{k}} \cdot \boldsymbol{\sigma},\\\nonumber
    G^{(h)}_0 \left(\mathbf{k}, i \omega_n \right) =& - G^{(+)} \left( -\mathbf{k}, -i\omega_n \right) \sigma_0 \\\label{HolePropagator}
    &+ G^{(-)} \left( -\mathbf{k}, -i\omega_n \right) \mathbf{\hat{g}}^{\mbox{\scriptsize eff}}_{\mathbf{-k}} \cdot \sigma_y \boldsymbol{\sigma} \sigma_y,
\end{align}
{where the unit vector $\mathbf{\hat{g}}_{\mathbf{k}}^{\text{eff}} \equiv \mathbf{g}_{\mathbf{k}}^{\text{eff}}/\abs{\mathbf{g}_{\mathbf{k}}^{\text{eff}}}$ is defined.}

We now calculate the second-order contribution ($n=1$) in the sum of Eq. \eqref{EffAcion}, i.e.,
\begin{align}\nonumber
    \delta \mathcal{F}^{(2)}=& \sum_{\mathbf{q}}\frac{\abs{\Delta_{\mathbf{q}}}^2}{g} \\
    &+ \frac{T}{2} \sum_{\mathbf{k}_1,\mathbf{k}_2,i\omega_n}\Tr{\mathcal{\hat{G}}_{0}(\mathbf{k}_1) \hat{\Delta} (\mathbf{k}_1,\mathbf{k}_2) \mathcal{\hat{G}}_{0}(\mathbf{k}_2) \hat{\Delta}(\mathbf{k}_2,\mathbf{k}_1)},
\end{align}
which yields
\begin{align}\nonumber
    \delta \mathcal{F}^{(2)} =& \Bigg(\frac{1}{g} - 2 T \sum_{\mathbf{k}, i \omega_n} \gamma^2(\mathbf{k}) \bigg\{  G^{(+)}  \widetilde{G}^{(+)} - G^{(-)}  \widetilde{G}^{(-)}\nonumber\\&\times\mathbf{{g}}^{\mbox{\scriptsize eff}}_{\mathbf{k}+\mathbf{Q/2}} \cdot \mathbf{{g}}^{\mbox{\scriptsize eff}}_{-\mathbf{k}+\mathbf{Q/2}}  \bigg\} \Bigg) \sum_{\mathbf{q}} \abs{\Delta_{\mathbf{q}}}^2\label{F2},
\end{align}
where we use the shorthand notation $ G^{(\pm)}= G^{(\pm)} \left( \mathbf{k}+{\mathbf{Q}}/{2}, i\omega_n \right)$ and $ \widetilde{G}^{(\pm)}= G^{(\pm)} \left( - \mathbf{k} +{\mathbf{Q}}/{2}, -i\omega_n \right)$.

Now, we introduce an energy cutoff $\epsilon_c$ by defining a range from $-\epsilon_c$ to $+\epsilon_c$ around the Fermi surface, which is the range of the pairing attraction, and change the integrals over momentum to integrals over energy and angle. This leads to the contribution $\delta \mathcal{F}^{(2)}$ given in Eq. \eqref{SecondOrderGL}, where we defined {$g_{1,2} = ( |{\mathbf{g}^{\text{eff}}_{\mathbf{k}+\mathbf{Q}/2}}| \pm |{\mathbf{g}^{\text{eff}}_{-\mathbf{k}+\mathbf{Q}/2}}| )/2$} and {$\cos  \eta_{\mathbf{k}}^{\mathbf{Q}}  = {\mathbf{\hat{g}}}^{\text{eff}}_{\mathbf{k}+\mathbf{Q}/2} \cdot {\mathbf{\hat{g}}}^{\text{eff}}_{-\mathbf{k}+\mathbf{Q}/2}$}. 
Therefore,
\begin{widetext}
    \begin{align}\nonumber
    \delta \mathcal{F}^{(2)} =& \Bigg[ \frac{1}{g} + \frac{N_F}{2} \sum_{\lambda =\pm 1} \int_{0}^{2\pi} \frac{d \theta}{2\pi} \gamma^2(\mathbf{k}) \Bigg(  \Bigg\{  \Xi \left( \frac{\lambda g_{1} + \frac{\mathbf{Q}}{2} \cdot \mathbf{v}}{2 \pi T}  \right) - \Xi \left( \frac{-i\epsilon_c + \lambda g_{1} + \frac{\mathbf{Q}}{2} \cdot \mathbf{v}}{2 \pi T}  \right)\Bigg\} \cos^2 \frac{\eta^{\mathbf{Q}}_{\mathbf{k}}}{2} + \Bigg\{ \Xi \left( \frac{\lambda g_{2} + \frac{\mathbf{Q}}{2} \cdot \mathbf{v}}{2 \pi T}  \right) \\\label{SecondOrderGL}
    & - \Xi \left( \frac{-i\epsilon_c + \lambda g_{2} + \frac{\mathbf{Q}}{2} \cdot \mathbf{v}}{2 \pi T}  \right) \Bigg\} \sin^2 \frac{\eta^{\mathbf{Q}}_{\mathbf{k}}}{2} \Bigg) \Bigg] \sum_{\mathbf{q} \in \{ \pm\mathbf{Q},\pm\mathbf{\widetilde{Q}} \}} \abs{\Delta_{\mathbf{q}}}^2 \equiv \mathit{a}(\mathbf{q}) \sum_{\mathbf{q} \in \{ \pm\mathbf{Q},\pm\mathbf{\widetilde{Q}} \}} \abs{\Delta_{\mathbf{q}}}^2.
\end{align}
The function $\Xi(x)$ is defined in terms of the digamma function $\psi^{(0)}(x)$ as $\Xi(x) = \Re[\psi^{(0)}\left(\frac{1}{2}+i x\right)]-\psi^{(0)}\left( \frac{1}{2} \right)$, and we considered a constant density of states in the vicinity of the non-interacting Fermi level given by $N_{F}$. Note that, in the limit of a large cutoff $\epsilon_c$, Eq. \eqref{SecondOrderGL} is in complete agreement with the expression obtained in Ref. \cite{Yuan_2022}. 

As for the fourth-order term ($n=2$), we have:
\begin{align}
    \delta \mathcal{F}^{(4)}=& \frac{T}{4} \sum_{\substack{\mathbf{k}_1,\mathbf{k}_2 \\ \mathbf{k}_3,\mathbf{k}_4, i\omega_n}}\Tr \left\{ \mathcal{\hat{G}}_{0}(\mathbf{k}_1) \hat{\Delta} (\mathbf{k}_1,\mathbf{k}_2) \mathcal{\hat{G}}_{0}(\mathbf{k}_2) \hat{\Delta}(\mathbf{k}_2,\mathbf{k}_3) \mathcal{\hat{G}}_{0}(\mathbf{k}_3) \hat{\Delta} (\mathbf{k}_3,\mathbf{k}_4) \mathcal{\hat{G}}_{0}(\mathbf{k}_4) \hat{\Delta}(\mathbf{k}_4,\mathbf{k}_1) \right\}.
\end{align}
Using the definitions of the free propagator and the order parameter matrices, we obtain
\begin{align}\nonumber
    \delta \mathcal{F}^{(4)}=& \frac{T}{2} \sum_{\mathbf{k}, i\omega_n} \sideset{}{'}\sum_{\substack{\mathbf{q_1},\mathbf{q_2}\\\mathbf{q_3} }} \Tr \Bigg\{ G_{0}^{(p)} \left( \mathbf{k}+\frac{\mathbf{q}_1}{2} \right) \sigma_y G_{0}^{(h)} \left( -\mathbf{k}+\frac{\mathbf{q}_1}{2} \right) \sigma_y G_{0}^{(p)} \left( \mathbf{k}-\frac{\mathbf{q}_1}{2} + \mathbf{q}_2 \right) \sigma_y G_{0}^{(h)} \left( -\mathbf{k}+\frac{\mathbf{q}_1}{2} - \mathbf{q}_2+\mathbf{q}_3 \right) \sigma_y \Bigg\}\\\label{FourthOrderGL}
     &\times \gamma\left( \mathbf{k} \right) \gamma\left( -\mathbf{k} + \frac{\mathbf{q}_1}{2} - \frac{\mathbf{q}_2}{2} \right) \gamma\left( \mathbf{k} - \frac{\mathbf{q}_1}{2} + \mathbf{q}_2 - \frac{\mathbf{q}_3}{2} \right) \gamma\left( -\mathbf{k} - \frac{\mathbf{q}_2}{2} + \frac{\mathbf{q}_3}{2} \right) \Delta_{\mathbf{q}_1} \Delta^{*}_{\mathbf{q}_2} \Delta_{\mathbf{q}_3} \Delta^{*}_{\mathbf{q}_1-\mathbf{q}_2+\mathbf{q}_3} \Bigg\}.
\end{align}
In Eq. \eqref{FourthOrderGL}, we have omitted for simplicity the dependence of the propagators on the Matsubara frequencies. The prime in the summation over the momenta $\mathbf{q}_i \in \{ \pm\mathbf{Q}, \pm \mathbf{\widetilde{Q}} \}$ ($i=1,2,3$) means that it has an additional constraint given by $\mathbf{q}_1-\mathbf{q}_2+\mathbf{q}_3 \in \{ \pm\mathbf{Q}, \pm \mathbf{\widetilde{Q}} \}$. {It is also worthwhile to emphasize here that the dependence of all GL coefficients on the microscopic parameters $t_{{AM}}, \alpha_R \text{ and } \mathbf{B}$ enter via the definition of the effective coupling $\mathbf{g}^{\text{eff}}_{\mathbf{k}}$ in Eqs. \eqref{ParticlePropagator} and \eqref{HolePropagator}.} As a result, we now explicitly show the contribution to each PDW phase described in detail in the main text:

a) \emph{Fulde-Ferrell} (FF) \emph{phase}: For this phase, we only select the order parameters with momentum $\mathbf{Q}$ in Eq. \eqref{FourthOrderGL}, {i.e.}, $\mathbf{q}_i = \mathbf{Q}$ (for all $i$), which gives:
\begin{align}\nonumber
    \delta \mathcal{F}^{(4)}_{FF}=& \frac{T}{2} \sum_{\mathbf{k}, i\omega_n} \gamma^4(\mathbf{k}) \Tr \Bigg\{ G_{0}^{(p)} \left( \mathbf{k}+\frac{\mathbf{Q}}{2} \right) \sigma_y G_{0}^{(h)} \left( -\mathbf{k}+\frac{\mathbf{Q}}{2} \right) \sigma_y G_{0}^{(p)} \left( \mathbf{k}+\frac{\mathbf{Q}}{2} \right) \sigma_y G_{0}^{(h)} \left( -\mathbf{k}+\frac{\mathbf{Q}}{2}\right) \sigma_y   \Bigg\} \abs{\Delta_{\mathbf{Q}}}^4\\\label{FourthOrderFF}
    \equiv & \chi_{1} \abs{\Delta_{\mathbf{Q}}}^4.
\end{align}    

b) \emph{Fulde-Ferrell} (FF$^*$) \emph{phase}: Here, there are two types of diagrams, {i.e.}, those containing four external legs with equal momenta $\mathbf{Q}$ or $\widetilde{\mathbf{Q}}$, and those with both momenta $\mathbf{Q}$ {and} $\widetilde{\mathbf{Q}}$. The former contributes with one diagram for each momentum with the same structure as $\chi_{1}$ in Eq. \eqref{FourthOrderFF}, while the latter results in four cases for the $\mathbf{q}_i$ in Eq. \eqref{FourthOrderGL}, namely $\mathbf{q}_1=\mathbf{Q}, \mathbf{q}_2=\mathbf{Q}, \mathbf{q}_3=\widetilde{\mathbf{Q}}, \mathbf{q}_4=\widetilde{\mathbf{Q}}$, and permutations of these. Using the fact that $G^{(h)}_0(\mathbf{k}, i\omega_n) = - \left[ G^{(p)}_0(-\mathbf{k}, -i\omega_n) \right]^{T}$ from Eqs. \eqref{ParticlePropagator} and \eqref{HolePropagator}, we can show that these diagrams are all equivalent and, after a change of variables, they yield:
\begin{align}\nonumber
    \delta \mathcal{F}^{(4)}_{FF^*}=& \chi_1 \left( \abs{\Delta_{\mathbf{Q}}}^4 + \abs{\Delta_{\widetilde{\mathbf{Q}}}}^4 \right) + 4\times \frac{T}{2} \sum_{\mathbf{k}, i\omega_n} \Tr \left\{ G_{0}^{(p)} \left( \mathbf{k}+\frac{\mathbf{Q}+ \widetilde{\mathbf{Q}}}{2} \right) \sigma_y G_{0}^{(h)} \left( -\mathbf{k}+\frac{\mathbf{Q}-\widetilde{\mathbf{Q}}}{2} \right) \sigma_y G_{0}^{(p)} \left( \mathbf{k}+\frac{\mathbf{Q}+\widetilde{\mathbf{Q}}}{2} \right) \sigma_y \right. \\
    &\left. \times G_{0}^{(h)} \left( -\mathbf{k}+\frac{-\mathbf{Q}+\widetilde{\mathbf{Q}}}{2} \right) \sigma_y \right\} \gamma\left( \mathbf{k} + \frac{\widetilde{\mathbf{\mathbf{Q}}}}{2} \right) \gamma\left( -\mathbf{k - \frac{\widetilde{\mathbf{Q}}}{2}} \right) \gamma\left( \mathbf{k} + \frac{\mathbf{Q}}{2} \right) \gamma\left( -\mathbf{k} - \frac{\widetilde{\mathbf{Q}}}{2} \right)  \abs{\Delta_{\mathbf{Q}}}^2 \abs{\Delta_{\widetilde{\mathbf{Q}}}}^2 \nonumber \\
    &\equiv \chi_1 \left( \abs{\Delta_{\mathbf{Q}}}^4 + \abs{\Delta_{\widetilde{\mathbf{Q}}}}^4 \right) + \chi_2 \abs{\Delta_{\mathbf{Q}}}^2 \abs{\Delta_{\widetilde{\mathbf{Q}}}}^2 .
\end{align}

c) \emph{Unidirectional} (UD) \emph{phase}: Here, we select order parameters with momentum $\mathbf{Q}$ and $-\mathbf{Q}$ in Eq. \eqref{FourthOrderGL}. There are again two types of diagrams: those with only one order parameter and the mixed terms. The latter result in four cases for the $\mathbf{q}_i$ in Eq. \eqref{FourthOrderGL}, namely $\mathbf{q}_1=\mathbf{Q}, \mathbf{q}_2=\mathbf{Q}, \mathbf{q}_3=-\mathbf{Q}, \mathbf{q}_4=-\mathbf{Q}$, and permutations of these, which are all equivalent using similar arguments as in the previous case. Rewriting them, we get:
\begin{align}\nonumber
    \delta \mathcal{F}^{(4)}_{UD}=& \chi_1 \left( \abs{\Delta_{\mathbf{Q}}}^4 + \abs{\Delta_{-\mathbf{Q}}}^4 \right) + 4\times \frac{T}{2} \sum_{\mathbf{k}, i\omega_n} \Tr \Bigg\{ G_{0}^{(p)} \left( \mathbf{k}+\frac{\mathbf{Q}}{2} \right) \sigma_y G_{0}^{(h)} \left( -\mathbf{k}+\frac{\mathbf{Q}}{2} \right) \sigma_y G_{0}^{(p)} \left( \mathbf{k}+\frac{\mathbf{Q}}{2} \right) \sigma_y \\
    & \times G_{0}^{(h)} \left( -\mathbf{k}-\frac{3\mathbf{Q}}{2} \right) \sigma_y \Bigg\} \gamma\left( \mathbf{k} \right) \gamma\left( -\mathbf{k} \right) \gamma\left( \mathbf{k} + \frac{\mathbf{Q}}{2} \right) \gamma\left( -\mathbf{k} - \frac{\mathbf{Q}}{2} \right) \abs{\Delta_{\mathbf{Q}}}^2 \abs{\Delta_{-\mathbf{Q}}}^2  \abs{\Delta_{\mathbf{Q}}}^2 \abs{\Delta_{-\mathbf{Q}}}^2\nonumber\\
    &\equiv \chi_1 \left( \abs{\Delta_{\mathbf{Q}}}^4 + \abs{\Delta_{-\mathbf{Q}}}^4 \right) + \chi_3 \abs{\Delta_{\mathbf{Q}}}^2 \abs{\Delta_{-\mathbf{Q}}}^2 .
\end{align}

d) \emph{Bidirectional} (BD) \emph{phases}: Now, there are three types of diagrams: those with only one order parameter, the mixed terms with absolute values of the order parameters and the mixed terms which involve the relative phase of the order parameters. The first and second types have already been calculated in the previous cases, with proper a change of external legs. The final result is then written as:
\begin{align}\nonumber
    \delta \mathcal{F}^{(4)}_{BD} =& \chi_1 \left(  \abs{\Delta_{\mathbf{Q}}}^4 + \abs{\Delta_{-\mathbf{Q}}}^4 +  \abs{\Delta_{\widetilde{\mathbf{Q}}}}^4 + \abs{\Delta_{-\widetilde{\mathbf{Q}}}}^4 \right) + \chi_2 \left( \abs{\Delta_{\mathbf{Q}}}^2 \abs{\Delta_{\widetilde{\mathbf{Q}}}}^2 + \abs{\Delta_{\mathbf{Q}}}^2 \abs{\Delta_{-\widetilde{\mathbf{Q}}}}^2 + \abs{\Delta_{-\mathbf{Q}}}^2 \abs{\Delta_{\widetilde{\mathbf{Q}}}}^2 + \abs{\Delta_{-\mathbf{Q}}}^2 \abs{\Delta_{-\widetilde{\mathbf{Q}}}}^2 \right)\\\nonumber
    &+\chi_3 \left( \abs{\Delta_{\mathbf{Q}}}^2 \abs{\Delta_{-\mathbf{Q}}}^2 + \abs{\Delta_{\widetilde{\mathbf{Q}}}}^2 \abs{\Delta_{-\widetilde{\mathbf{Q}}}}^2 \right) + 4\times \frac{T}{2} \sum_{\mathbf{k}, i\omega_n} \Tr \Bigg\{ G_{0}^{(p)} \left( \mathbf{k}+\frac{\mathbf{Q}+\widetilde{\mathbf{Q}}}{2} \right) \sigma_y G_{0}^{(h)} \left( -\mathbf{k}+\frac{\mathbf{Q}-\widetilde{\mathbf{Q}}}{2} \right) \sigma_y \\
    &G_{0}^{(p)} \left( \mathbf{k}+\frac{-\mathbf{Q}-\widetilde{\mathbf{Q}}}{2} \right) \sigma_y G_{0}^{(h)} \left( -\mathbf{k}+\frac{-\mathbf{Q}+\widetilde{\mathbf{Q}}}{2} \right) \sigma_y \Bigg\} \gamma\left( \mathbf{k} + \frac{\widetilde{\mathbf{Q}}}{2} \right) \gamma\left( -\mathbf{k} + \frac{\mathbf{Q}}{2} \right) \gamma\left( \mathbf{k} - \frac{\widetilde{\mathbf{Q}}}{2} \right) \gamma\left( -\mathbf{k} - \frac{\mathbf{Q}}{2} \right) \nonumber\\
    &\times\left( \Delta_{\mathbf{Q}} \Delta_{-\mathbf{Q}} \Delta^{*}_{\widetilde{\mathbf{Q}}} \Delta^{*}_{-\widetilde{\mathbf{Q}}} + \Delta^{*}_{\mathbf{Q}} \Delta^{*}_{-\mathbf{Q}} \Delta_{\widetilde{\mathbf{Q}}} \Delta_{-\widetilde{\mathbf{Q}}}  \right).\label{F4_BD}
    \end{align}    
\end{widetext}
From the above expressions and by defining the last term of Eq. \eqref{F4_BD} as the coefficient $\chi_4$, we obtain the general form of the GL free energy shown in Eq. \eqref{general_F} in the main text. Note that this last contribution to the free energy becomes $+2\chi_4$ for the BD1 phase, and a similar term becomes instead $-2\chi_4$ for the BD2 phase. All fourth-order coefficients of the GL expansion are calculated numerically here. In this way, we compute the above sums within the interval determined by the energy cutoff defined by $\epsilon_c=10$ meV and for the physical parameters given by $m =0.005$ meV$^{-1}$ and $T=10$ K.

\text

%

\end{document}